# New β-Pyrochlore Oxide Superconductor $CsOs_2O_6$

Shigeki Yonezawa*, Yuji Muraoka and Zenji Hiroi

*Institute for Solid State Physics, University of Tokyo, Kashiwa, Chiba 277-8581*

**Abstract**

The discovery of a new β-pyrochlore oxide superconductor $CsOs_2O_6$ with $T_c$ = 3.3 K is reported. It is the third superconductor in the family of β-pyrochlore oxides, following $KOs_2O_6$ with $T_c$ = 9.6 K and $RbOs_2O_6$ with $T_c$ = 6.3 K. The $T_c$ of this series decreases with increasing the ionic radius of alkaline metal ions, imposing negative chemical pressure upon the Os pyrochlore lattice.

Pyrochlore oxides constitute a large family of transition metal (TM) oxides like perovskites.[1] They have the general chemical formula $A_2B_2O_7$ or $A_2B_2O_6O$', where A is a larger cation and B is a smaller TM cation. The first superconductor in the family of pyrochlore oxides was discovered in $Cd_2Re_2O_7$ at $T_c$ = 1.0 K.[2-4] Recently, we found another type of pyrochlore oxides with the general formula $AB_2O_6$ called the β-pyrochlore oxide,[5] where A is a large monovalent alkaline metal cation. Two osmates, $RbOs_2O_6$ and $KOs_2O_6$, have been prepared in this family, which exhibit superconductivity at higher $T_c$'s of 6.3 K and 9.6 K, respectively.[5,6] They crystallize in a modified pyrochlore structure, where Os atoms form a corner-sharing tetrahedral network called the pyrochlore lattice, as in $A_2B_2O_7$-type pyrochlore oxides which is now called α-pyrochlore, while alkaline metal atoms occupy the 8b site which is the O' site in the α-pyrochlore. Here we report on the discovery of another new β-pyrochlore oxide superconductor $CsOs_2O_6$.

A polycrystalline sample was prepared using conventional solid-state reaction method in an evacuated silica tube. To control the oxygen partial pressure, a certain amount of AgO was added separately from the pellet in the tube: AgO decomposes into silver and oxygen above 370 K, and thus generates an oxidizing atmosphere. The chemical composition of the product examined by the energy dispersive X-ray (EDX) analysis in a scanning electron microscope was Cs:Os ~ 1:2.

Figure 1 shows a powder X-ray diffraction (XRD) pattern taken at room temperature. All the intense peaks can be indexed assuming a cubic unit cell with a lattice constant $a = 1.0149$ nm. A few extra peaks from Os are detected. Moreover, a trace of unknown impurity phase is also included in the product. Extinctions observed in the XRD pattern are consistent with the space group of $Fd\bar{3}m$, and the intensity profile is similar to the other β-pyrochlore oxides. The lattice constant is larger than those of $RbOs_2O_6$ ($a = 1.0114$ nm) and $KOs_2O_6$ ($a = 1.0101$ nm), as expected from the difference in the ionic radius of alkaline metal ions.

Resistivity measurements were carried out down to 2 K by the standard four-probe method in a Quantum Design PPMS. Figure 2 shows the temperature dependence of resistivity measured on a polycrystalline sample. It exhibits good metallic behavior below room temperature. A clear $T^2$ dependence is seen below 45 K. As shown in the inset to Fig. 2, the resistivity shows a sharp drop below 3.4 K due to superconductivity. The zero resistivity is attained below 3.2 K. The critical temperature $T_c$ defined as the midpoint temperature of the transition is 3.3 K.

In addition to the observation of the zero-resistive transition, a large diamagnetic signal associated with the Meissner effect was observed below 3.3 K. Figure 3 shows the temperature dependence of magnetic susceptibility measured on a powdered sample in a Quantum Design MPMS. The measurements were carried out in a magnetic field of 10 Oe on heating after zero-field cooling and then on cooling in a field. A superconducting volume fraction estimated at 2 K from the zero-field cooling experiment is nearly 100 %, indicating bulk superconductivity.

Now we have three β-pyrochlore oxide superconductors; $CsOs_2O_6$, $RbOs_2O_6$ and $KOs_2O_6$. The $T_c$ changes with alkaline metals as 3.2 K, 6.3 K and 9.6 K, respectively. It is plausible to assume that this change is due to the size effect of alkaline metal ions, because the bands near the Fermi level consist of Os $5d$ orbitals with minor contribution from O $2p$ orbitals in the case of $Cd_2Os_2O_7$[7] and probably in $KOs_2O_6$, too.[8] Certainly, the lattice constant is almost proportional to the ionic radius of A ions. The relation between the $T_c$ and lattice constant is shown in Fig. 4. The $T_c$ decreases with increasing $a$ under negative chemical pressure. This is in contrast to the case of conventional BCS superconductivity in a single band model, where the $T_c$ may increase under negative pressure, because the density of state (DOS) increases. The reverse tendency found in $AB_2O_6$ may partly reflect the complex band structure with many sharp peaks in the DOS. However, it can be also related to the mechanism of superconductivity, which would be clarified by systematic study on these compounds. It is expected from the figure that positive pressure would raise the $T_c$.   High pressure experiments are now in progress.

In conclusion, we found superconductivity with $T_c = 3.3$ K in the new β-pyrochlore oxide $CsOs_2O_6$.   Although the nature of this superconductivity is not known at the moment, we believe that an interesting aspect of physics is involved in the superconductivity of $CsOs_2O_6$, as in $KOs_2O_6$ and $RbOs_2O_6$.[9]

We thank F. Sakai and M. Matsushita for their help in the EDX analysis and H. Ueda for his helpful advice on synthesis.   We also thank J. Yamaura and T. Muramatsu for fruitful discussion.   This research was supported by a Grant-in-Aid for Scientific Research on Priority Areas (A) provided by the Ministry of Education, Culture, Sports, Science and Technology, Japan.

**References**
1) M. A. Subramanian, G. Aravamudan and G. V. Subba Rao: Prog. Solid State Chem. **15** (1983) 55.
2) M. Hanawa, Y. Muraoka, T. Tayama, T. Sakakibara, J. Yamaura and Z. Hiroi: Phys. Rev. Lett. **87** (2001) 187001.
3) H. Sakai, K. Yoshimura, H. Ohno, H. Kato, S. Kambe, R. E. Walstedt, T. D. Matsuda, Y. Haga and Y. Onuki: J. Phys.: Condens. Matter **13** (2001) L785.
4) R. Jin, J. He, S. McCall, C. S. Alexander, F. Drymiotis and D. Mandrus: Phys. Rev. B **64** (2001) 180503.
5) S. Yonezawa, Y. Muraoka, Y. Matsushita and Z. Hiroi: J. Phys. Soc. Jpn. **73** (2004) 819.
6) S. Yonezawa, Y. Muraoka, Y. Matsushita and Z. Hiroi: J. Phys.: Condens. Matter **16** (2004) L9.
7) D. J. Singh: Phys. Rev. B **65** (2002) 155109.
8) H. Harima: private communication
9) Z. Hiroi, S. Yonezawa and Y. Muraoka: submitted to J. Phys. Soc Jpn.

**Figure captions**

Fig. 1.   Powder X-ray diffraction pattern of $CsOs_2O_6$.   Peak index is given by assuming a cubic unit cell with a lattice constant $a = 1.0149$ nm.   Asterisks mark extra peaks from Os.

Fig. 2.   Temperature dependences of resistivity measured on a polycrystalline sample. Inset shows enlargements around the superconducting transition.

Fig. 3.   Temperature dependence of magnetic susceptibility measured on a powdered sample of $CsOs_2O_6$ in an applied field of 10 Oe.   ZFC and FC indicate zero-field cooling and field cooling curves, respectively.

Fig. 4.   Relation between the lattice constant $a$ and critical temperature $T_c$ for three β-pyrochlore oxide superconductors; $CsOs_2O_6$, $RbOs_2O_6$ and $KOs_2O_6$.   $Cd_2Os_2O_7$ is an α-pyrochlore osmate which is a non-superconductor.

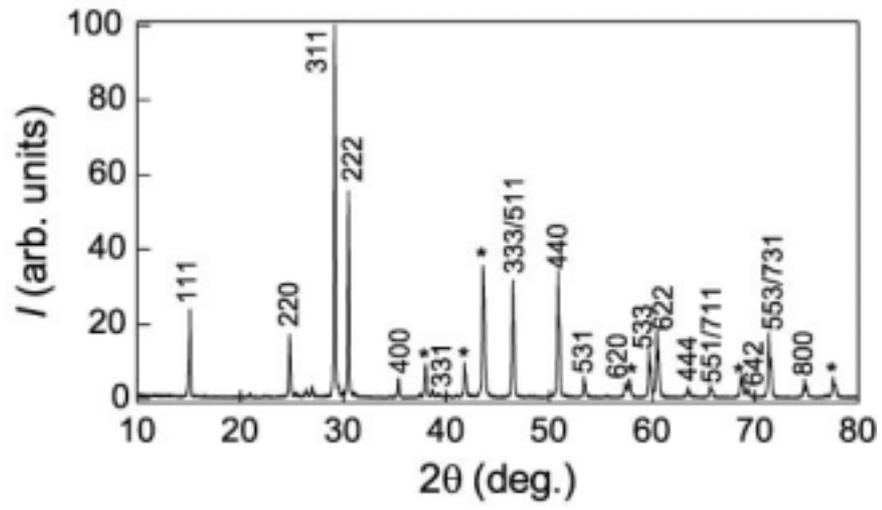

Fig. 1.

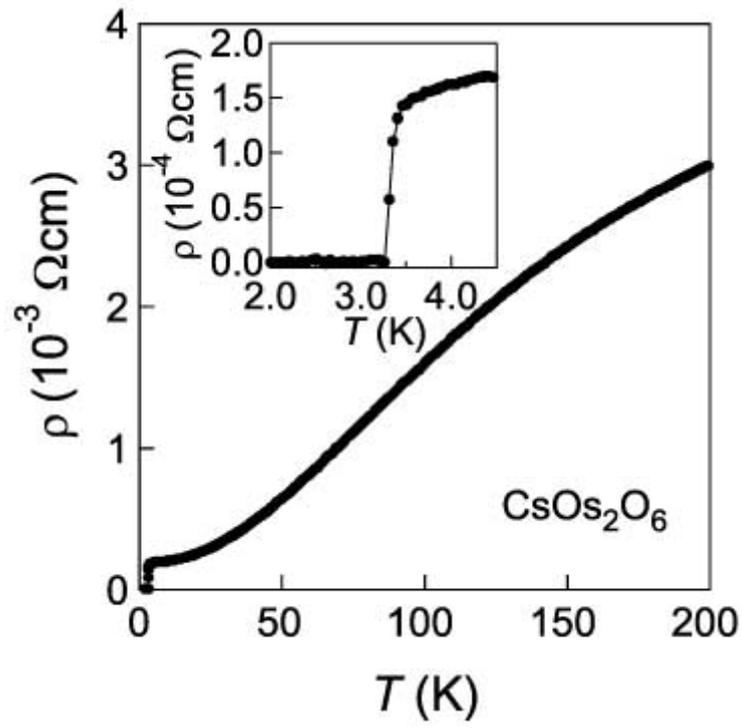

Fig. 2.

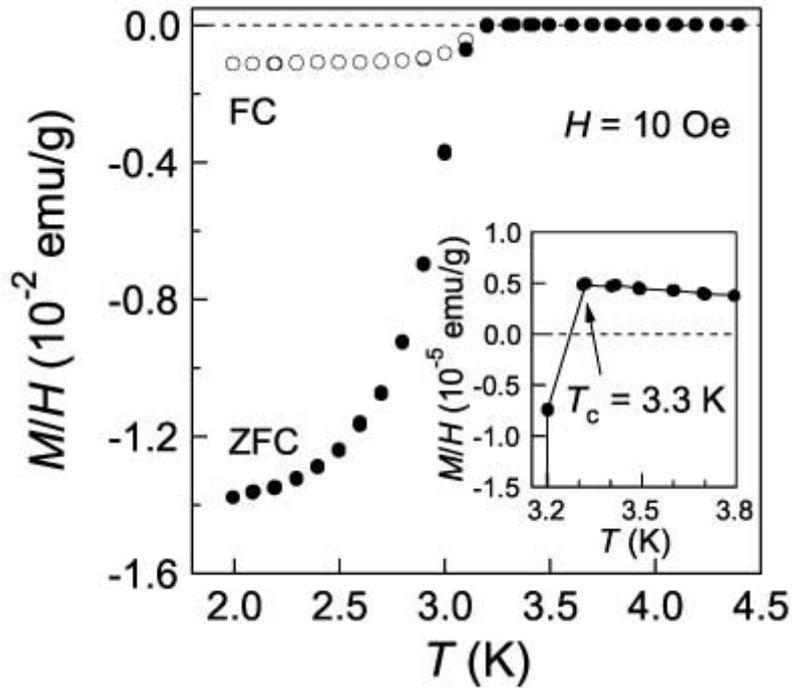

Fig. 3.

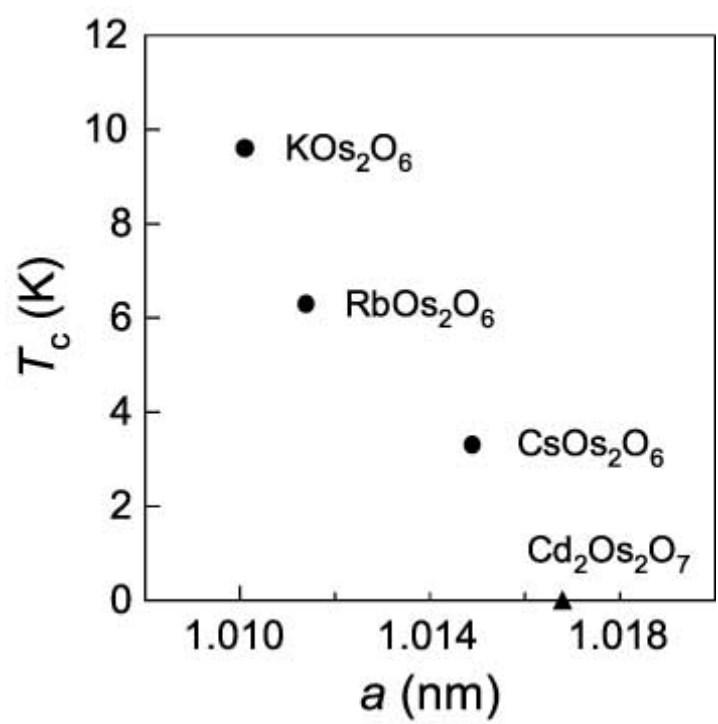

Fig. 4.